\documentclass[5p,preprint]{elsarticle}

\usepackage{lineno,hyperref}
\usepackage{amsmath}
\usepackage{hyperref}
\usepackage{graphicx}
\usepackage{amssymb}
\usepackage[utf8]{inputenc}
\usepackage[normalem]{ulem}

\journal{Chaos Solitons \& Fractals}

\biboptions{sort&compress}
\begin{document}

\begin{frontmatter}

\title{An epidemiological model with voluntary quarantine strategies governed by evolutionary game dynamics}

\author[1]{Marco A. Amaral \fnref{corr}}
\address[1]{Instituto de Artes, Humanidades e Ciências, Universidade Federal do Sul da Bahia, Teixeira de Freitas-BA,  45996-108 Brazil}
\fntext[corr]{Corresponding author: marcoantonio.amaral@gmail.com}

\author[2]{Marcelo M. de Oliveira }
\address[2]{Departamento de F\'isica e Matem\'atica, CAP, Universidade Federal de S\~ao Jo\~ao del Rei, Ouro Branco-MG, 36420-000 Brazil}

\author[3]{Marco A. Javarone }
\address[3]{University College London, London, UK}




\begin{abstract}
During pandemic events, strategies such as social distancing can be fundamental to reduce simultaneous infections and mitigate the disease spreading, which is very relevant to the risk of a healthcare system collapse. Although these strategies can be recommended, or even imposed, their actual implementation may depend on the population perception of the risks associated with a potential infection. The current COVID-19 crisis, for instance, is showing that some individuals are much more prone than others to remain isolated. To better understand these dynamics, we propose an epidemiological SIR model that uses evolutionary game theory for combining in a single process social strategies, individual risk perception, and viral spreading. In particular, we consider a disease spreading through a population, whose agents can choose between self-isolation and a lifestyle careless of any epidemic risk. The strategy adoption is individual and depends on the perceived disease risk compared to the quarantine cost. The game payoff governs the strategy adoption, while the epidemic process governs the agent's health state. At the same time, the infection rate depends on the agent's strategy while the perceived disease risk depends on the fraction of infected agents. Our results show recurrent infection waves, which are usually seen in previous historic epidemic scenarios with voluntary quarantine. In particular, such waves re-occur as the population reduces disease awareness. Notably, the risk perception is found to be fundamental for controlling the magnitude of the infection peak, while the final infection size is mainly dictated by the infection rates. Low awareness leads to a single and strong infection peak, while a greater disease risk leads to shorter, although more frequent, peaks. The proposed model spontaneously captures relevant aspects of a pandemic event, highlighting the fundamental role of social strategies. 
\end{abstract}


\end{frontmatter}


\section{Introduction}
\label{intro}
During a pandemic, quarantine and other distancing rules can constitute the only option to curb the viral spreading, in particular in absence of vaccines or medicines to control the symptoms resulting from an infection ~\cite{Chua2010, Manfredi2013, Ali2020, Keeling2011}.
Usually, these social rules are defined by epidemiologists and other experts, however their actual implementation can be quite challenging. For instance, the current COVID-19 crisis ~\cite{hopkins_univ_website, Tagliazucchi2020, Bavel2020} is showing how some people are more easily prone to self-isolate under voluntary quarantine than others, even despite evidences on the potential risks. By doing so, individuals that avoid any form of restriction become an element of risk for themselves and for their community.
In these scenarios, understanding how to stimulate and sustain prosocial behaviors has a paramount relevance. In this work, we aim to study the relationship between human behavior, represented by individual quarantining strategies, and the epidemic spreading of a disease.
We emphasize that this model is not an empirical description of the current COVID-19 evolution. Instead, this is a general theoretical framework that merges evolutionary game theory (EGT) ~\cite{Szabo2007} and epidemiology in a single compartmental model. Such framework allows rational strategy changes between agents and can be used to better understand the central aspects regarding a generic epidemic event. Note that this model focus on voluntary self-imposed quarantines only. Such strategy is different from policy-driven quarantines, where the government can apply contact tracing and other methods to enforce the isolation of specific individuals that were shown to be infected, or potentially infectious \cite{Braithwaite2020, Aleta2020, MorenoLopez2020}.

Usually, the approach for studying a pandemic or epidemic process is based on compartmental models~\cite{Keeling2011, Brauer2012, Zhang2020}, which are a ubiquitous tool in epidemiology and modern health management systems. 
The SIR model is one of the most known epidemiological models ~\cite{Keeling2011, Brauer2012, Chang2020}. It describes the spreading of a disease, which confers immunity against re-infection, in agents that evolve from the susceptible compartment, $S$, to the infectious, $I$, and eventually to the recovered (or removed) compartment $R$. Although simple, it has been widely used to obtain relevant aspects of epidemic processes that present the $S \rightarrow I \rightarrow R$ structure. 
Since its introduction in the seminal paper by Kermack and McKendrick ~\cite{Kermack-McKendrick1927}, the model has been extensively studied and expanded to consider different hypotheses and conditions. 
For example, some epidemics may require more compartments, such as the exposed and/or asymptomatic agents (known as SEIR and SEAIR models respectively) ~\cite{Keeling2011, Anderson-May,Daley-Gani, Kaddar2011}. 
Spread on complex networks was also proven useful to understand the heterogeneity of agent contacts ~\cite{Cota2018, Sander2013, pastor-satorras_rmp15, Mata2015, Shaw2008}. 
The study of control and mitigation strategies such as vaccination ~\cite{Wang2016a}, modeling of vector-borne diseases ~\cite{Pinho2010, DeSouza2013}, and effects of birth-and-death dynamics ~\cite{Manfredi2013, Brauer2012} are other examples of the wide range of applications for compartmental models in epidemiology. 
Even rumors and corruption spreading have found a natural framework in the SIR model ~\cite{Bauza2020,Lu2020a, Askarizadeh2019, Zhao2013, Amaral2020b, Amaral2018b, DeArruda2016}.
Nevertheless, most of those models relate only to the disease evolution, i.e. agents usually have no conscious actions regarding the disease. 
On the other hand, many control measures for infectious diseases depend on individual decision making. In this context, the recent field of behavioral epidemiology is attracting the attention of researchers from diverse areas, applying psychology, social engineering, and game theory approaches to epidemiology (see~\cite{Manfredi2013, Chang2020, Verelst2016} for a review). Instead of considering agents having static roles, behavioral epidemiology includes dynamic behavior changes.
This is a fertile ground for the recent area of social dynamics, or sociophysics ~\cite{galam_ijmpc08, Capraro2018, Perc2017}, which utilizes tools from statistical physics together with evolutionary game theory (and others) to better understand the complex behavior of humans ~\cite{Bavel2020, Perc2016, Sigaki2018,  DOrsogna2015, castellano_rmp09, Javarone2016c, javarone_book, Kumar2020}.
For example, in a novel approach, Bauch ~\cite{Bauch2003, Bauch2005, bauch_pnas04} integrated a SIR model into an EGT framework to analyze vaccination decision dynamics. 
By doing so, agents change their vaccination strategy dynamically, depending on their perception of the benefits and costs of a vaccine.
This was later generalized into the so-called `vaccination games' framework (see ~\cite{Wang2016a} for a comprehensive review).
Such approach led to many interesting observations and predictions in vaccination protocols ~\cite{bauch_pnas04, Arefin2020, Chang2020, Wang2020, Kabir2019a, Kabir2019, Kazuki2018, Iwamura2018, Ferrer-i-cancho2012, d-onofrio_jtb11, Basu2008, Ye2020, RajibArefin2019a, Alam2020, Jentsch2020}. %

Recent works also investigated other mitigation strategies such as awareness campaigns ~\cite{Steinegger2020}, wealth differences ~\cite{Shuler2019, Kordonis2020}, economic incentives ~\cite{Kabir2020}, social distancing ~\cite{Chowdhury2020, reluga_ploscb10}, information spreading ~\cite{Kabir2019b, DaSilva2019}, multi-layer contact networks ~\cite{Kabir2019c}, dynamic contacts ~\cite{Zanette2008} and others ~\cite{Rowlett2020, Poletti2009, lagorio_pre11, Biswas2020, Arenas2020a}. A general overview of these investigations shows the presence of a cycle, where effective mitigation measures lead to a low risk perception, which in turn weakens said mitigation strategies, bringing the disease back ~\cite{Manfredi2013}. The most recent anti-vaccination movement is just one of a long history of such cycles ~\cite{Gangarosa1998, Pearce2008, Luman2004}. 
Unfortunately, vaccination is not always an option, and social isolation can be the only practice to prevent further disease spread ~\cite{kato_sr11, Chua2010, Manfredi2013, Arenas2020}.
Such was the case in the famous episodes of the Spanish flu ~\cite{Morens2007, Flores2018}, SARS epidemic of 2002–2003 ~\cite{Feng2009, CHAN-YEUNG2003} and more recently, during the COVID-19 pandemic ~\cite{Sohrabi2020, Jiang2020, Ali2020, Tagliazucchi2020, Ndairou2020, Zhang2020a, Arenas2020a}. 

In the present work, we propose a ``\textit{quarantine game}'', in which agents undergo a SIR epidemic process while, at the same time, they can choose between two actions, i.e. to self-quarantine and voluntarily stay at home (Q), or continue acting normally (N). Following the game theory usual nomenclature, here we use the word strategy as meaning the agent's chosen action (quarantine or not). The strategy is constantly updated based on the individual perceived cost of the quarantine versus the perceived disease risk.
While the scope of the model is intentionally general, it is mainly motivated by the recent COVID-19 global pandemic and its consequences, that have shown a wide spectrum of human responses to the viral spreading. For instance, countries adopted many different restriction policies, from mild distancing rules to strict lock-down. 
However, when not mandatory, only a small fraction of individuals may decide to self-isolate, while the rest of a community avoids restrictions, endangering themselves and others. 
The fast scale of this phenomenon has also shown how collective perceptions of the disease risk has changed in a matter of weeks (based or not on real scientific data) ~\cite{Altmann2020}. 
This can be seen from how individuals and policymakers across the world have so far considered a variety of options, spanning from strict lock-downs to doing nothing, with the hope of reaching some kind of herd immunity ~\cite{Kay2020, Studdert2020, Randolph2020}.
The variety of social strategies adopted worldwide, and in particular their results in terms of successes and failures, constitute a relevant evidence of how important is the behavioral component of a given strategy during pandemic events.

Lastly, we emphasise that this is a theoretical model, and in no way intends to fully grasp all the social and political complexities exhibited by the current pandemic scenario \cite{Castro2020}. 
On the contrary, it aims to merge two elements of paramount relevance in these scenarios, i.e. game theory and epidemic spreading, on a singular time scale.

\section{Model} 
\label{model}

\begin{figure}
\centering
  \includegraphics[width=8cm]{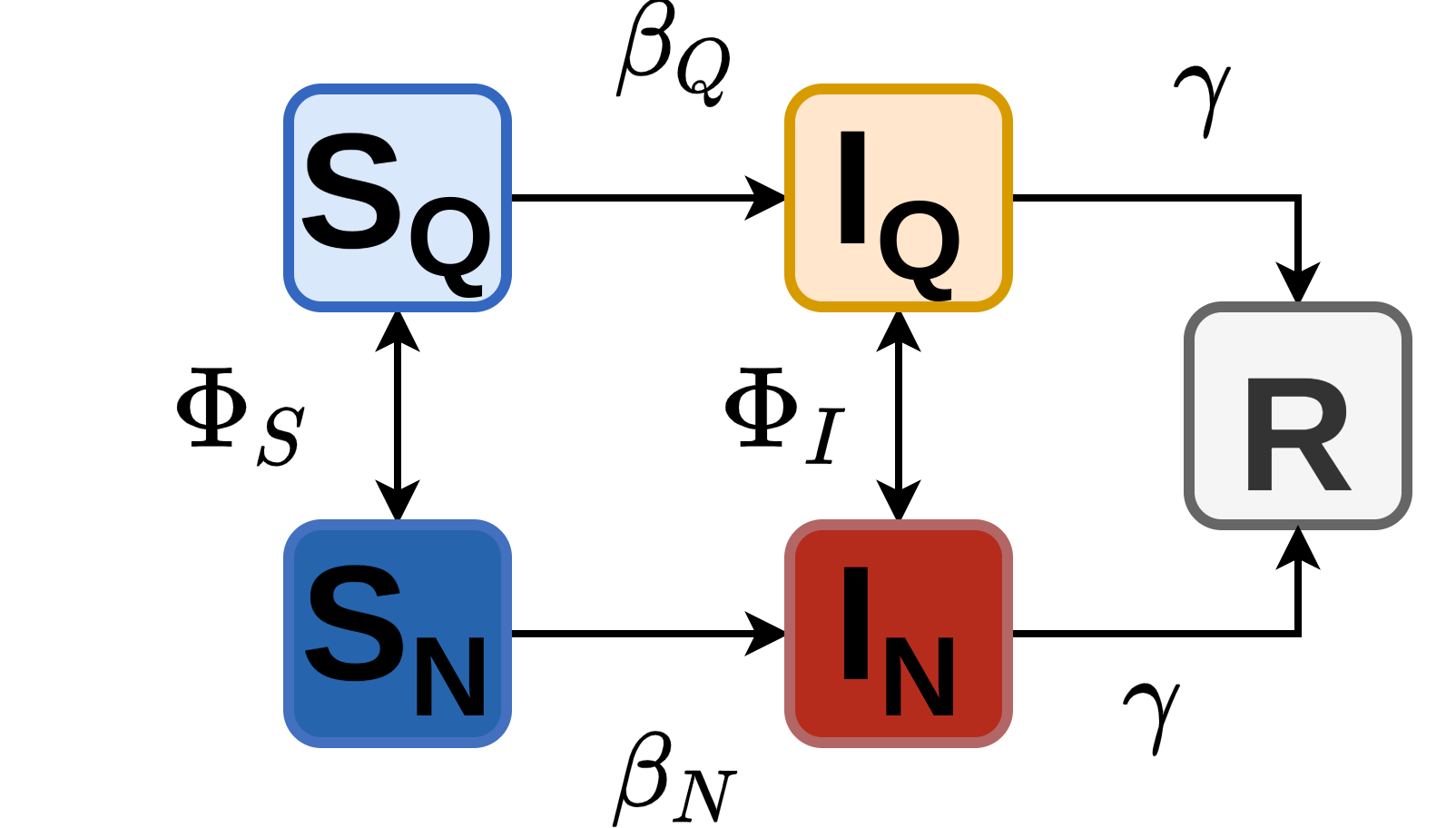}
  \caption{Schematic representation of the proposed model. We consider five compartments where agents transition from $S$, $I$, and $R$ states through epidemiological dynamics. At the same time, agents change their own strategy ($Q$ or $N$) through an evolutionary game dynamics. The parameter $\beta_i$ is the infection rate that, depending on the strategy of an agent, is defined as $\beta_Q$ or $\beta_N$ (i.e. quarantine versus normal life stile). Also note that infected individuals of one given strategy can interact with susceptible individuals of the other strategy, e.g. $S_Q-I_N$, by the cross-infection rate $\beta_a$, further explained in the text. The parameter $\gamma$ represents the recovery rate and is independent of the specific strategy. $\Phi$ represents the strategy change flux for each epidemic state and it is governed by the evolutionary game dynamics.}
  \label{diagram}
\end{figure}

In the proposed model, susceptible agents (S) become infected (I) with a rate $\beta_i$ upon contact with another infected agent. Then, at a constant rate $\gamma$, infected agents get recovered (R). 
Besides, agents can self-impose a quarantine (Q) and stay at home, or keep acting as in a normal situation (N). In the language of game theory, the former strategy, can be interpreted as a form of cooperation, while the latter as a selfish behavior, i.e. a form of defection. Therefore, we shall refer interchangeably to agents adopting quarantine as cooperators, and agents acting normally, as defectors. 
We note that in reality, the epidemiological term \textit{quarantine} is only applicable to someone that is not infected and chooses to stay at home. If someone is infected and chooses this action, the correct term would be \textit{isolation}. Nevertheless, for simplicity, here we name this strategy as self-quarantine for both susceptible and infected agents, as such differentiation in the equations would lead only to a more complex nomenclature.
The main effect of the chosen action is to influence the individual infection rate $\beta_i$. We assume that quarantined agents have a lower infection
rate than normal ones, that is $\beta_Q<\beta_N$, since those agents reduce their interactions with other members of their community.
Also, note that although rare, cross-interactions between the two types of strategies (quarantine or not) can still occur in our model, e.g. $S_Q$ becomes infected by interacting with an $I_N$ individual. The cross-infection rate $\beta_a$ is used in such scenarios and we expect that, in general, $\beta_Q<\beta_a<\beta_N$. Deep explanation of such parameter is given in the following paragraphs.
We expand the usual SIR model into a five compartment model, $S_Q$, $S_N$, $I_Q$ $I_N$, and $R$. As recovered agents cannot be infected again, their chosen strategy is irrelevant. An illustrative diagram is shown in Figure ~\ref{diagram}. 

By using a compartmental approach, the evolutionary game dynamics is fully integrated into the model. This differs from usual behavioral epidemiology approaches where the strategy fraction evolves according to a separate dynamic  ~\cite{Rowlett2020, Poletti2009, Bauch2005, Wang2016a, Chowdhury2020, reluga_ploscb10, Manfredi2013, d-onofrio_jtb11}. 
Hence, our model allows cross interactions (such as $S_Q$ interacting with $I_N$), giving rise to a rich scenario where sub-population correlations can be observed.

Employing the game theory concept of {\em perceived payoff} ($\pi$), agents base their strategies on the perceived risk of their current action. A cooperator (i.e. an agent self-imposing quarantine) expects to suffer a perceived cost $\Omega$. This represents the difficulties one might face in a period of quarantine, but in turn, it strongly reduces the probability of being infected. This leads to a constant payoff (or perceived risk) for cooperators, $Q$:

\begin{equation}\label{cpay}
\pi_Q=-\Omega.
\end{equation}

On the other hand, defectors, i.e. agents adopting the strategy $N$, have a perceived risk based on their infection probability multiplied by the perceived disease cost parameter $\delta$: 

\begin{equation}\label{dpay}
\pi_N=-\delta \beta_N I.
\end{equation}

We remark that the payoffs are based exclusively on the agent's individual perceptions. This is in accordance with the widespread notion of individual risk perception based on the number of (anecdotal) cases an agent is exposed to ~\cite{Fu2011, Palekar2008, Johnson1983, DaSilva2019}. The game theory dynamics concerns what agents perceive to be their risks and rewards, and not necessarily the actual risk of a given action.
It is also interesting to understand how the model considers infection sub-notifications, i.e. general population perception of the fraction $I$ being lower than the actual level. By using a linear payoff structure, sub-notifications can be absorbed in a re-scaled $\delta'$ value. E.g., if the informed infected fraction of the population is sub-notified by a fraction $f$, the payoff structure would be the same, while the re-scaled perceived risk would just  be $\delta'= \delta (1-f)$. Note that the sub-notification only affects the payoff function, while the epidemiological dynamics still depends only on the actual fraction of infected individuals.

Following the usual evolutionary game dynamics, the probability of a given agent $i$ to adopt the strategy of agent $j$ is related to their payoffs $\pi_i$ and $\pi_j$. We use the typical Fermi rule ~\cite{Szabo2007}:
\begin{equation}\label{fermi}
\Theta(\pi_i,\pi_j)= \frac{1}{1+e^{-(\pi_j-\pi_i)/k}}.
\end{equation}
This allows strategy revision with a small but non zero chance of mistakes. Such irrationality is measured by the $k$ parameter, set as $k=0.1$ ~\cite{Szabo2007, Perc2017, Javarone2016d}. 
To obtain the total fraction of agents changing to a given strategy at any moment, we consider the number of encounters between any kind of $Q$ and $N$ strategies, inside each health compartment ($S$ or $I$), and multiply it by the strategy transition probability $\Theta(\pi_i,\pi_j)$ between strategies $i$ and/or $j$. This is equivalent to the master equation (for each compartment) of an evolutionary game dynamic ~\cite{Szabo2007, Amaral2018} using the mean-field approximation, and leads us to the strategy conversion rates, defined as
\small
\begin{align}\label{phisimpleS}
 \Phi_{S}=S_Q(S_N+I_N)\Theta(\pi_Q,\pi_N)-S_N(S_Q+I_Q)\Theta(\pi_N,\pi_Q) \\ \label{phisimpleI}
 \Phi_{I}=I_Q(S_N+I_N)\Theta(\pi_Q,\pi_N)-I_N(S_Q+I_Q)\Theta(\pi_N,\pi_Q).
\end{align}
\normalsize
Here, $\Phi_{S}$ is the rate at which $S_Q$ agents convert to $S_N$ (and conversely for $\Phi_{I}$), and it is governed by the EGT part of the model. 

Regarding the infection dynamics, we assume three different infection rates, that is, $\beta_N>\beta_a>\beta_Q$. Here, $\beta_N$ is the infection rate for defectors interacting with defectors, and similarly, $\beta_Q$ is the infection rate for cooperators. Cooperators and defectors interact through the cross-infection rate $\beta_a$. For the sake of simplicity we set $\beta_a=a(\beta_N+\beta_Q)/2$, an average value of $\beta_Q$ and $\beta_N$ weighted by the external control parameter $1>a>0$. We set $a=0.1$ to allow a small but non zero chance of cross-infection. The recovery rate is assumed to be the same for all agents. 
Considering all the assumptions above, we present the equations that describe the proposed model,

\begin{align}\label{SIRCD}
  \dot{S_N}&=-S_N(\beta_N I_N+\beta_a I_Q)+\tau\Phi_S \\
  \dot{S_Q}&=-S_Q(\beta_a I_N+\beta_Q I_Q)-\tau\Phi_S \\
  \dot{I_N}&=S_N(\beta_N I_N+\beta_a I_Q)-\gamma I_N+\tau\Phi_I \\
  \dot{I_Q}&=S_Q(\beta_a I_N+\beta_Q I_Q)-\gamma I_Q-\tau\Phi_I \\
  \dot{R}&=\gamma(I_N+I_Q),
\end{align}

where $\tau$ is the coupling parameter that controls how quickly one adopts a new strategy, in relation to the time-scale of the epidemic. Note that the current version of the model does not include vital dynamics, such as birth and death processes, since the model focuses on spread dynamics that take place in a matter of months.

\section{Results} \label{results}

We start by noting that the payoff structure proposed in Eqs. (\ref{cpay}) and (\ref{dpay}), is akin to the public goods and climate change dilemma games ~\cite{ Wang2019, Wardil2017, Gois2019, Yang2020} where each agent payoff depends on the total number of agents in some other state. That is, the quarantine game is not a pairwise interaction game such as the prisoner dilemma ~\cite{Szabo2007}. In particular, in our case, the defector payoff depends on the total number of infected agents ($I$), either cooperators or defectors, while the cooperator payoff is constant.
In doing so, we obtain the collective equivalent of the snow-drift game (also known as chicken or hawk-dove game ~\cite{Szabo2007}), i.e. as long as most of the population is healthy (susceptible or recovered), the best strategy is to defect and to continue acting normally. But as soon as most of the population chooses this strategy, the amount of infected agents grows, resulting in a change of the best strategy, that becomes to self-quarantine. 
It is also worth mentioning that such scenario is akin to the minority-game (or El Farol Bar dilemma ) \cite{arthur_aer94, Biswas2020, challet_04}, where each single individual receives the optimal payoff if she chooses the least chosen strategy on average.
Such payoff structure can be seen as a general anti-coordination game class, where the best strategy is to do the opposite of what your opponents are doing. Or specifically in our case, the opposite of what the majority of the population is doing ~\cite{Szabo2007}. 
However, note that the fraction of infected agents is not equal to the fraction of defectors, due to the epidemiology dynamics.  This is similar to the dilemma presented in vaccination games ~\cite{Bauch2003, Bauch2005, Iwamura2018, d-onofrio_jtb11, Fu2011} where agents should vaccinate but, as long as the majority of the population is vaccinated, the incentive to not vaccinate grows. This anti-coordination element is a central driver for the observed oscillatory dynamics.

The numerical integration of the equations is obtained through a 4th order Runge–Kutta method. For the interested reader, a simplified Python script for solving the equations is available at~\cite{github}. 
Regarding the results, unless stated otherwise, we set $\Omega=1,\tau=1,\gamma=1,\beta_Q=1, k=0.1, a=0.1$ and focus on the effects of varying the infection risk perception, $\delta$, and defectors infection rate $\beta_N$. 
As initial condition, the starting setting for the population has only a very small fraction of infected agents, i.e. $I_0=0.01, S_0=1-I_0$., while strategies are equally divided between $C$ and $D$.

\begin{figure}
\centering
  \includegraphics[width=8cm]{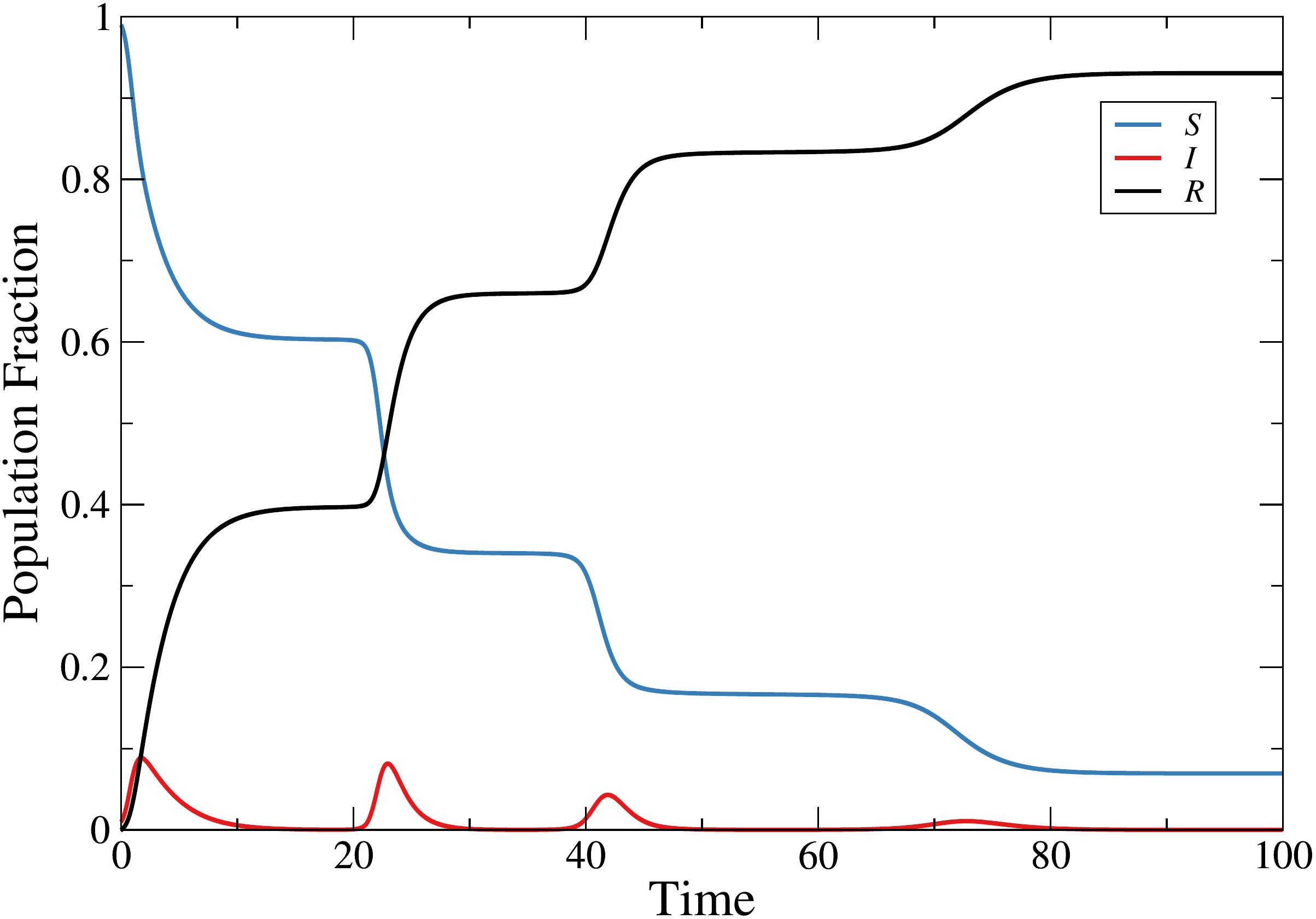}
  \caption{Typical behavior of the epidemiological population, $S=(S_Q+S_N), I=(I_Q+I_N), R$. Note that recurrent infection peaks emerge spontaneously.  Here $\delta=10, \beta_N=10$.}
  \label{typical}
\end{figure}

Figure \ref{typical} presents the typical behavior of the population. The most evident phenomenon is the recurrent infection waves, even though the model has no explicit oscillatory terms.  Looking at the evolution of the epidemiological population, i.e. $S=(S_Q+S_N)$, $I=(I_Q+I_N)$, and $R$, we notice that susceptible agents diminish on almost discrete steps. The successive drops in $S$ also coincide with the peaks of infected agents.
The inclusion of voluntary quarantine procedures in the SIR model spontaneously generates recurrent infection periods. This phenomenon can be observed for a wide range of parameters and it is a characteristic behavior of the model. 
Note that such an effect is similar to the expected scenario of real quarantine policies~\cite{Ali2020, Sohrabi2020, Manfredi2013}, that is, re-occurring infection seasons. Interestingly, previous pandemics as the Spanish flu (1918) presented such infection wave behavior~\cite{Chowell2006, Chowell2006a}.

\begin{figure}
\centering
  \includegraphics[width=8cm]{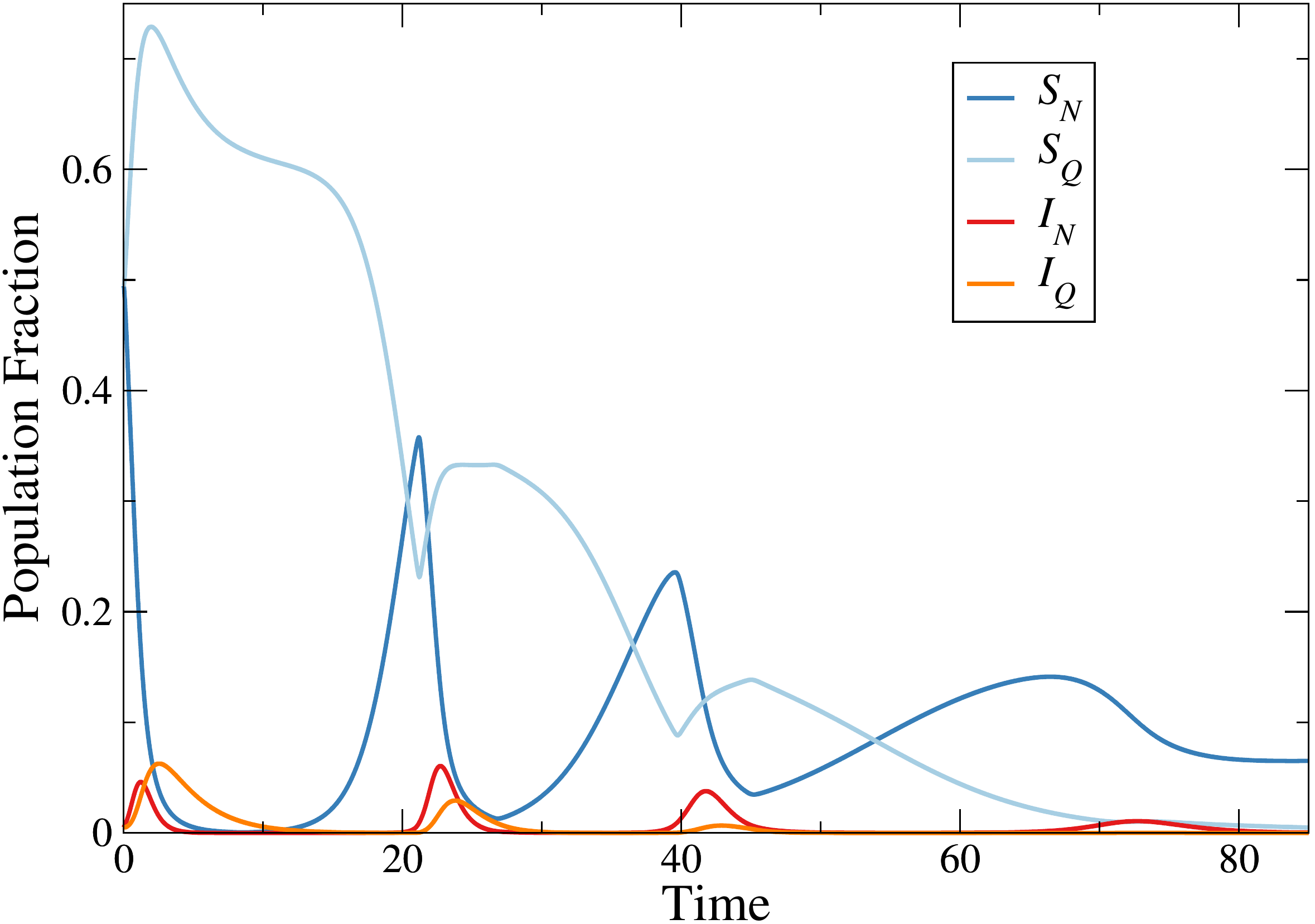}
  \caption{Typical behavior of the sub-population, $S_Q, S_N, I_Q, I_N, R$. The successive infection peaks are due to the frequent oscillations in the strategies, even if the total susceptible and removed individuals do not oscillate. Here, $\delta=10, \beta_N=10$.}
  \label{subpopevo}
\end{figure}

The cause underlying the successive infection peaks can be understood looking at the sub-population ($S_Q, S_N, I_Q, I_N, R$) and the strategy distributions through time. This can be seen in Figure \ref{subpopevo}. 
Remarkably, the population behavior hides a complex dynamic. In particular, as the fraction of infected agents initially grows, the cooperator's payoff quickly becomes advantageous. This is what causes the first broad peak of $S_Q$, as most agents start to undergo quarantine. In turn, the total fraction of infected agents begin to decline, as the majority of the population gets quarantined, with a low value of infection rate.
Nevertheless, as $I$ tends to $0$, the payoff for agents leaving quarantine (defector strategy) starts to grow and eventually it becomes greater than the cooperator's payoff. This triggers a flux of $S_Q\rightarrow S_N$, that is, people leaving quarantine. Such an event corresponds to the sharp increase in $S_N$, near the beginning of the second infection wave. 
With more and more agents leaving quarantine, a second peak of infected agents inevitably occurs. Indeed, we see that the infection peaks are always preceded by a sharp increase in the defector density. At this point, $S_N$ begins to decrease sharply because part of them becomes infected and the others (still susceptible) start becoming cooperators (the second and broad peak in $S_Q$). 
This process repeats itself again and again, at each time with less active agents.
An interesting effect also occurs in the sub-population of infected agents, i.e. the infection peak on defectors always precedes the peak of cooperators. We note that the number and height of the peaks, and recurrent infection cycles, are highly dependent on $\delta$.

Next, we analyze the mixed strategy equilibrium point to obtain the strategy inflection points. This is a similar approach as the one used in ~\cite{bauch_pnas04} for vaccination games. Suppose a mixed strategy where an agent has a probability $P$ to cooperate. This leads to the average expected payoff of $\bar{\pi}=P\pi_Q+(1-P)\pi_N$. We want to maximize it in relation to $P$, therefore:

\begin{equation}
\bar{\pi}=P(\delta \beta_N I-\Omega)-\delta \beta_N I .
\end{equation}

Since all parameters are greater than zero, we obtain the maximum expected payoff value when $P=1$ (always cooperate) if $\delta \beta_N I>\Omega$. Conversely, if $\delta \beta_N I<\Omega$, the maximum average payoff occurs for $P=0$ (always defect). This implies that agents will start changing strategies at an infection level of:

\begin{equation}\label{inflexpt}
I'=\frac{\Omega}{\delta\beta_N}
\end{equation}

\begin{figure}
\centering
  \includegraphics[width=8cm]{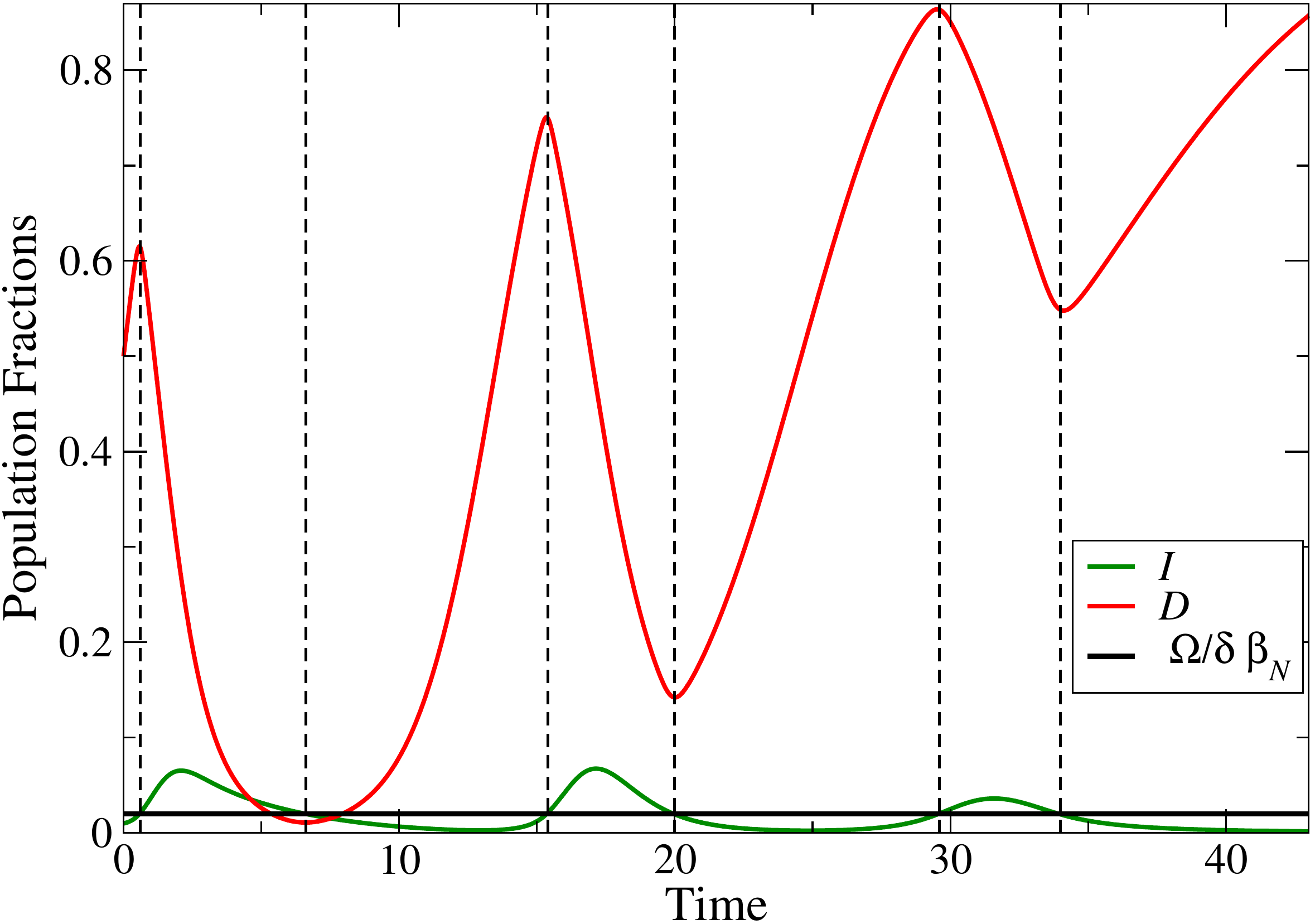}
  \caption{Infected agents ($I$) and defectors strategy fraction ($D$) time evolution. The horizontal line represents the value $I'=\Omega/ \delta \beta_N$. The vertical dashed lines indicate when $I(t)=I'$. As expected, these are the strategy maximum and minimum values. Here $\delta=10, \beta_N=5$.}
  \label{inflectionfig}
\end{figure}

In a system composed of fully rational agents, the strategy maximum and minimum values will coincide with the points mentioned above. Numerical analysis of the ODE integration shows good agreement with such prediction even if we use the Fermi strategy probability (an approach that has inherent fluctuations/irrationality). This can be seen in Figure \ref{inflectionfig}. We note that the main effect of greater irrationality, i.e. larger values of $k$, is to make the strategy oscillations more smooth around the inflection points. This analysis remained accurate for all studied values of $\delta, \Omega$, and $\beta_N$. Also, note that the peak of infections always happens between a maximum and minimum value of $D$, in a way consistent with all studied values of parameters.

\begin{figure*}
\centering
  \includegraphics[width=5.9cm]{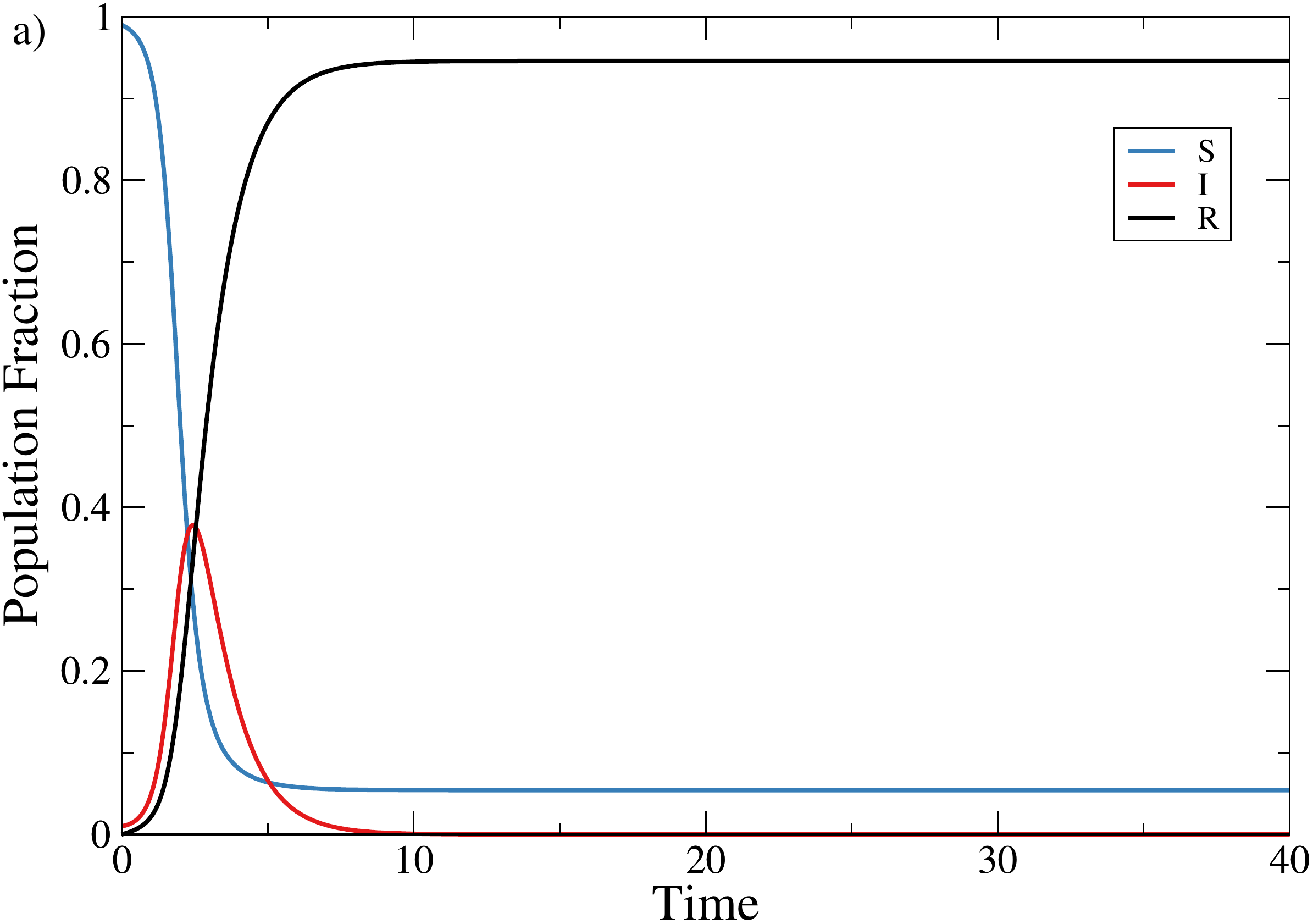}
  \includegraphics[width=5.9cm]{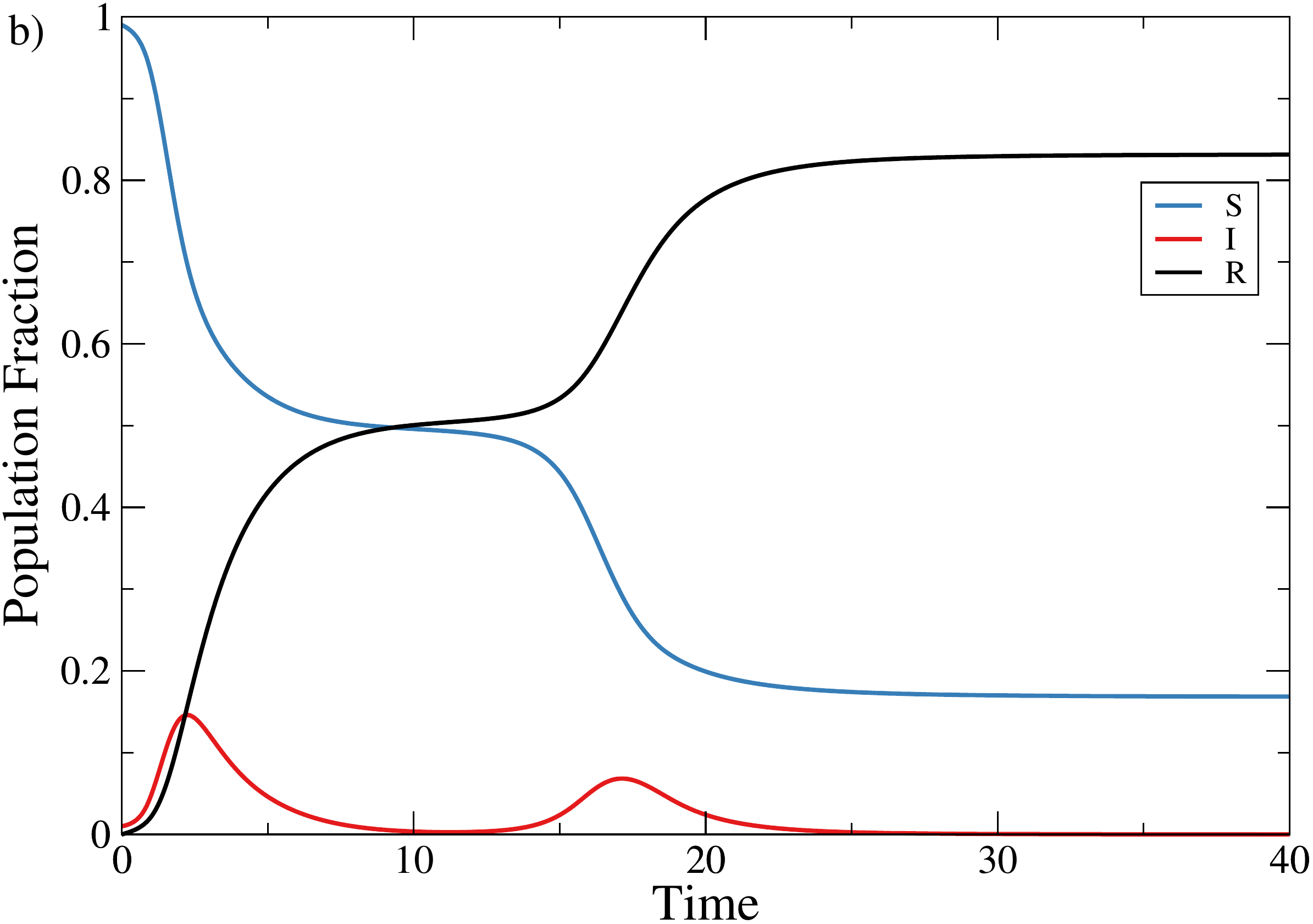}
  \includegraphics[width=5.9cm]{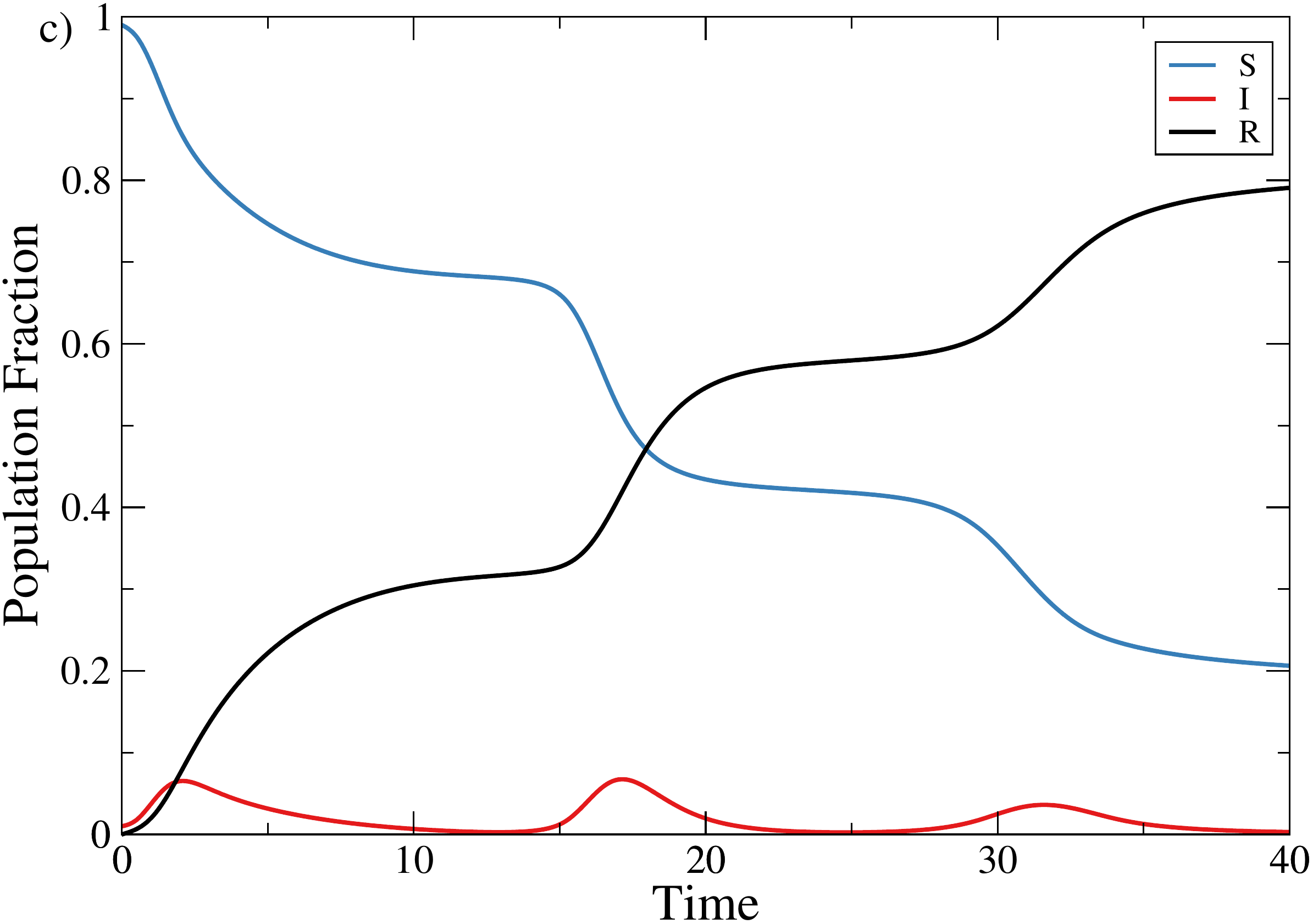}
  \caption{Typical behavior for diverse disease risk perception. In a) there is no disease risk perception, $\delta=0$, and the disease behaves according to the usual SIR dynamics, with a big and singular infection peak. In b) $\delta=5$ and while there are two infection waves, their magnitude is considerably smaller. Finally, in c) $\delta=10$, and we can see three shallow infection peaks. Note that as $\delta$ increases, the infection are distributed during a longer time span. In general, an increase in risk perception leads to smaller, and more distributed, infection peaks. Here $\beta_N=5$.}
  \label{dvar}
\end{figure*}

To better understand the effect of the disease risk perceptions on the infection peak size and duration, we vary the value of $\delta$, as this is the central parameter we are interested in. Figure \ref{dvar} shows the population dynamics when $\delta=\{0;~5;~10\}$. For low-risk perceptions, agents leave quarantine earlier and in great numbers. This creates a big single infection peak, which is consistent with the current worst-case scenarios for a pandemic ~\cite{ Ali2020, Sohrabi2020, Manfredi2013}. As we increase the risk perception, agents will tend to cooperate (stay in quarantine) for longer periods, leading to the distribution of smaller infection peaks along one or more infection cycles. We highlight that this is an emergent behavior that spontaneously appears by considering the evolutionary game dynamics.
In a pandemic scenario, this can be one of the most important aspects of a quarantine policy, since the healthcare system may have a small capacity, and cannot take care of all infected agents at the same time ~\cite{Zhang2020}.

\begin{figure}
\centering
  \includegraphics[width=8cm]{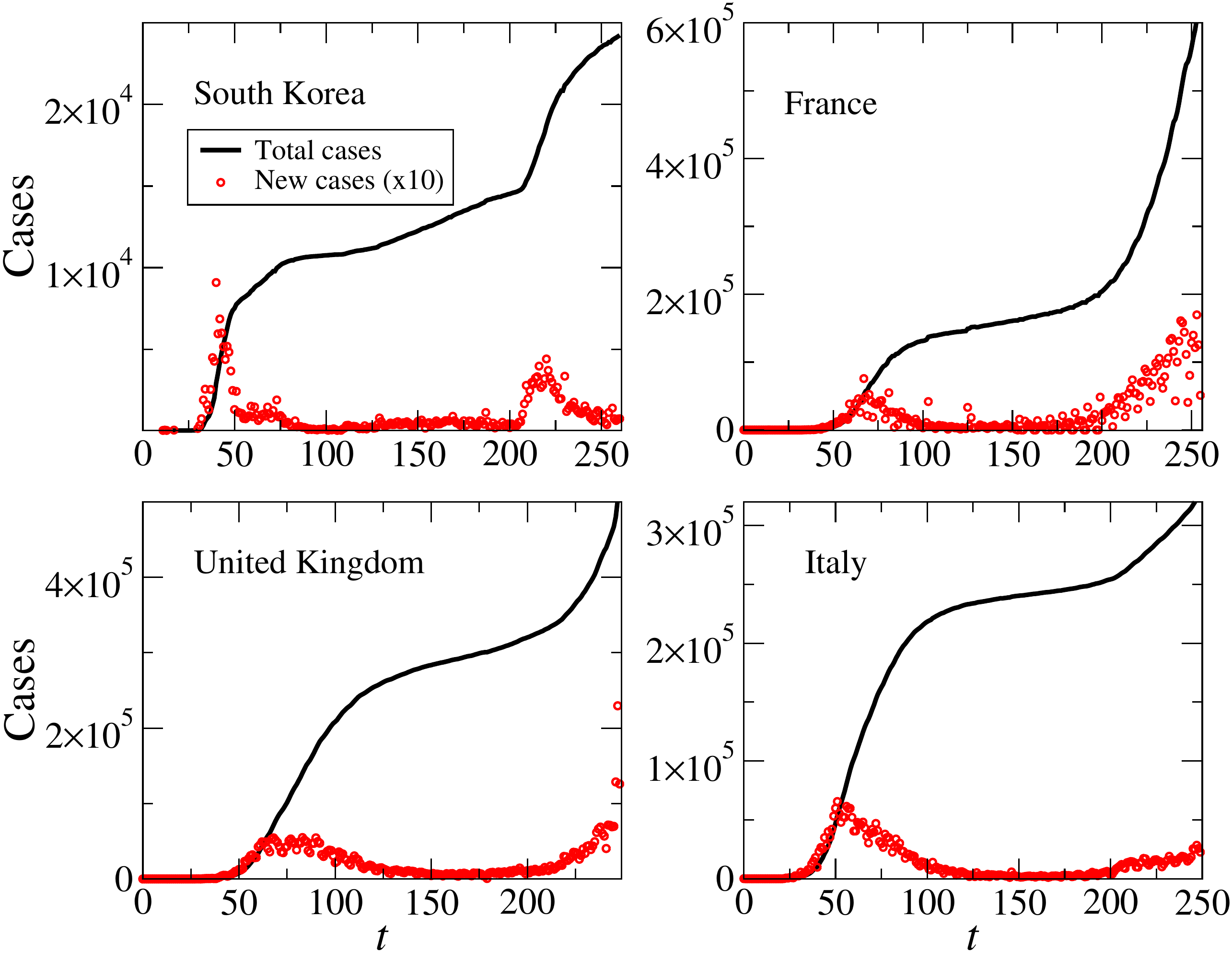}
\caption{Real data from four different countries regarding the number of total and new cases ($\times 10$) since the beginning of the epidemic. While the presented model do not aim to be an empirical fit, it is remarkable to see how the general behavior of secondary infection waves is present. Data obtained from \cite{data_world}.}
  \label{data}
\end{figure}

The central characteristic of the model is the spontaneous emergence of recurrent infection waves during an epidemic without the possibility of re-infection (SIR). While this is not an empirical model, it is insightful to look for similar general patterns in real data from the current COVID-19 crisis. Figure ~\ref{data} presents the actual data (obtained from \cite{data_world}) of four different countries, regarding the reported number of new cases ($\times 10$) and the total number of infected individuals. Such numbers are equivalent to the fractions of $I$ and $R$ in our model respectively. It is possible to see a remarkable similarity in the general behaviour of the model with the presented data regarding the infection peaks, as well as the stair-like increase in the total number of cases. Nevertheless, we stress that here we are not trying to fit the real data to our model, just observe how the general behaviour present similarities. This can also be seen in data from other countries, as the ones presented in \cite{Cacciapaglia2020} and from the major data banks like \cite{data_world}.

\begin{figure}[h] 
\centering
  \includegraphics[width=8cm]{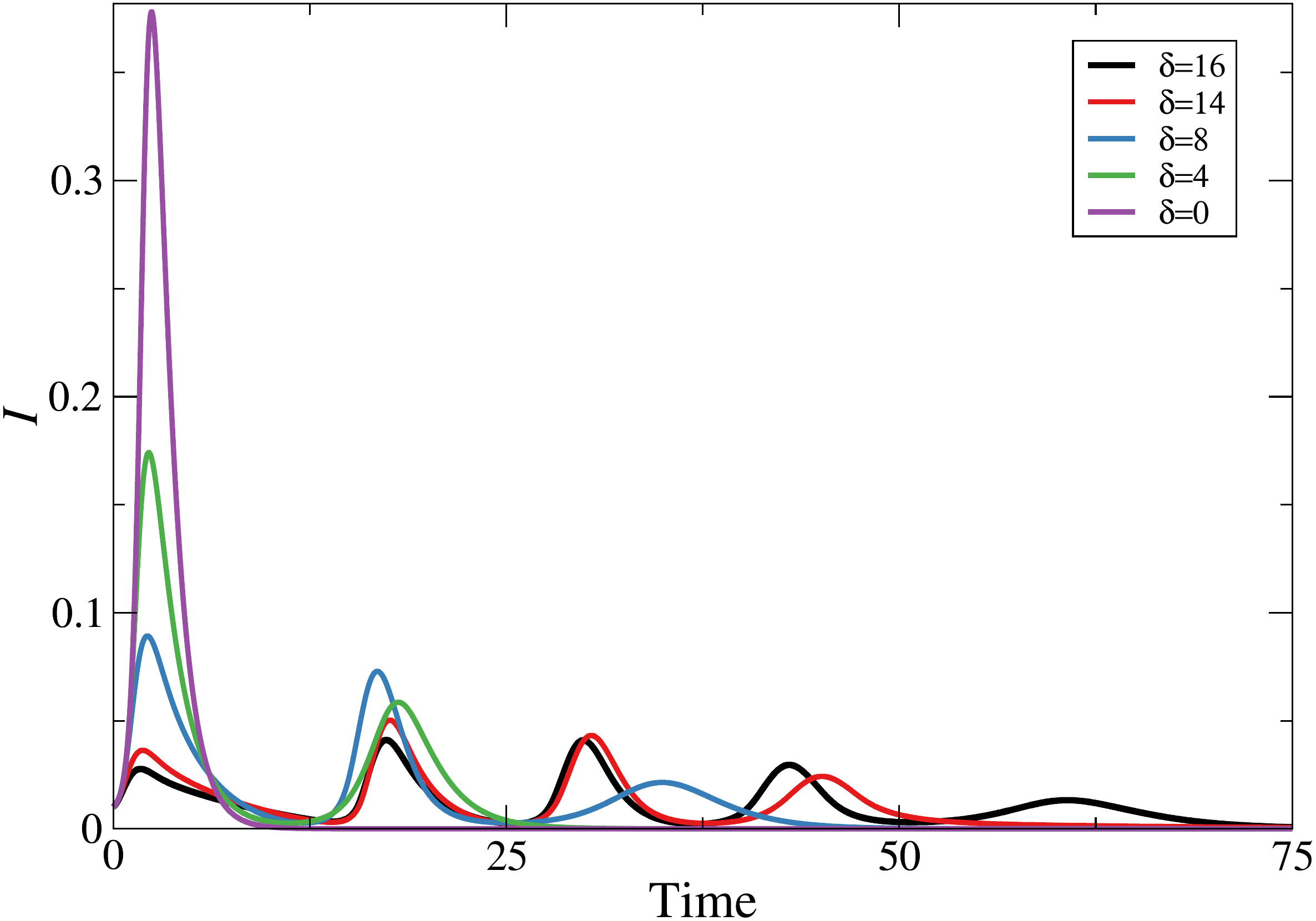}
  \caption{ Evolution of infected agents for different disease perception values, $\delta$. This parameter plays a key role in the infection peak magnitude, making it shorter while distributing the cases over many smaller infection waves. Here $\beta_N=5$.}
  \label{vardelta2}
\end{figure}

The effects of different disease perception values are summarized in Figure ~\ref{vardelta2}. Note that when $\delta=16$ there are even five different infection peaks, all with a very small magnitude. Another interesting effect to observe is that the first infection peak is not always the highest. For larger values of $\delta$, the highest peak can happen after some initial (small) infection wave.  Moreover, a higher risk perception better distributes the cases over long periods.

\begin{figure}
\centering
  \includegraphics[width=8cm]{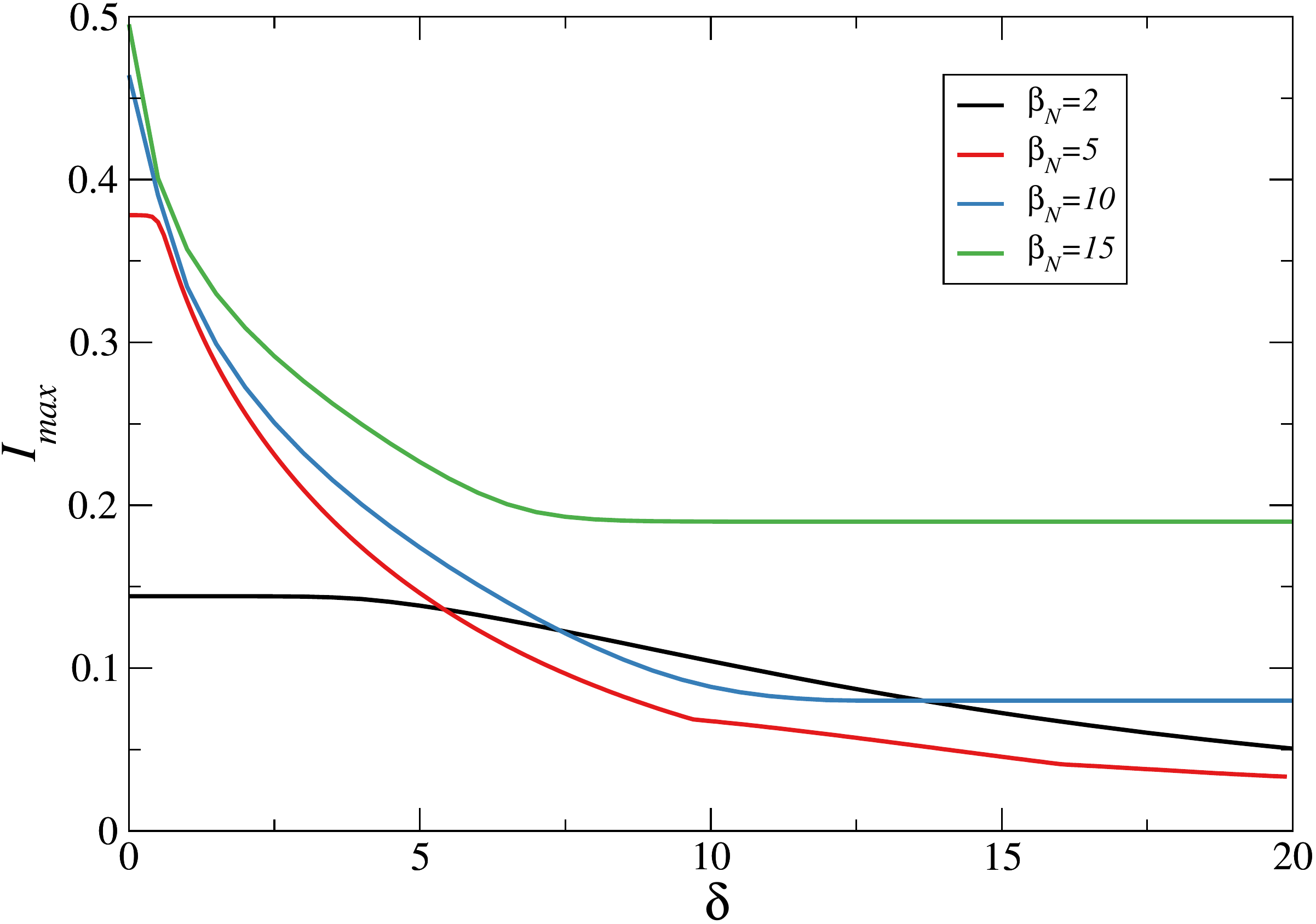}
  \caption{Maximum simultaneous infected agents density ($I_{max}$) as a function of the perceived disease risk $\delta$. The magnitude of the peak decreases for greater disease awareness values.}
  \label{peakheight}
\end{figure}

We emphasize that the infection peak magnitude can be a very important quantity when dealing with pandemics ~\cite{Zhang2020}. In Figure \ref{peakheight} we present the maximum simultaneous infection size ($I_{max}$) as a function of the perceived disease risk for different defector infection rates, $\beta_N$. We highlight that the equations of the proposed model can always be normalized in relation to $\beta_Q$, defining a new time scale. 
Because of this, without loss of generality, we chose to vary only $\beta_N$ in the presented results. We see that the disease awareness, $\delta$, can greatly help diminish the maximum simultaneous infected number. On the other hand, the effect of $\beta_N$ in $I_{max}$ is less pronounced. 

\begin{figure}
\centering
  \includegraphics[width=8cm]{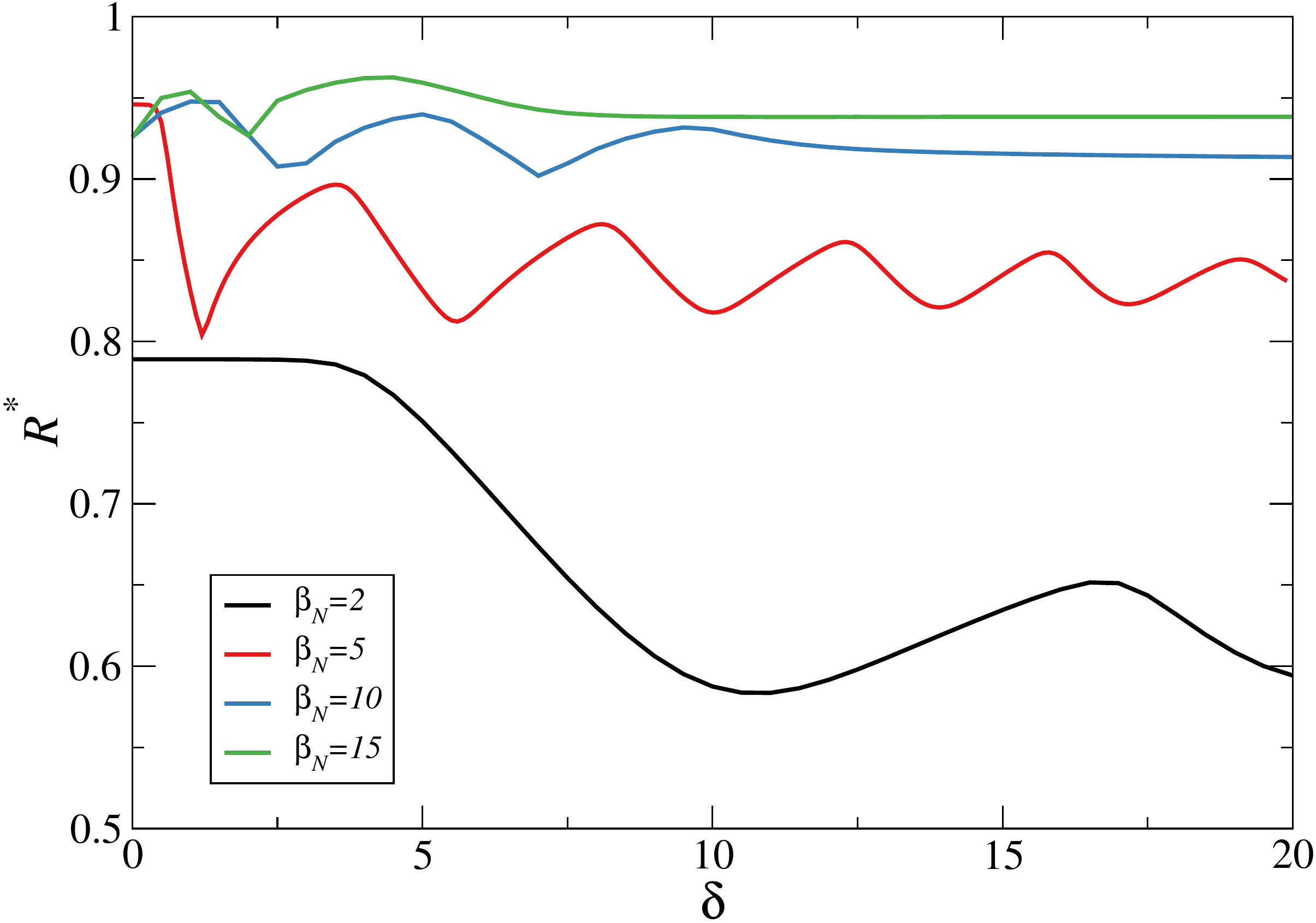}
  \caption{Infection size, measured as the final value of removed agents, $R^*$. The impact of $\delta$ in $R^*$ is less pronounced than in $I_{max}$. For some values of $\beta_N$, the fraction $R^*$ presents oscillations.}
  \label{infectionsize}
\end{figure}

We now analyze the infection size, measured by the final density of removed agents, $R^*$, shown in Figure ~\ref{infectionsize}. We note that the increase in $\delta$ can lead, on average, to slightly smaller $R^*$ values. The decrease is more pronounced when $\beta_N<2$. Differently from $I_{max}$ however, the behavior of $R^*$ is not monotonous in $\delta$, presenting non-periodic oscillations.

\begin{figure}
\centering
   \includegraphics[width=8cm]{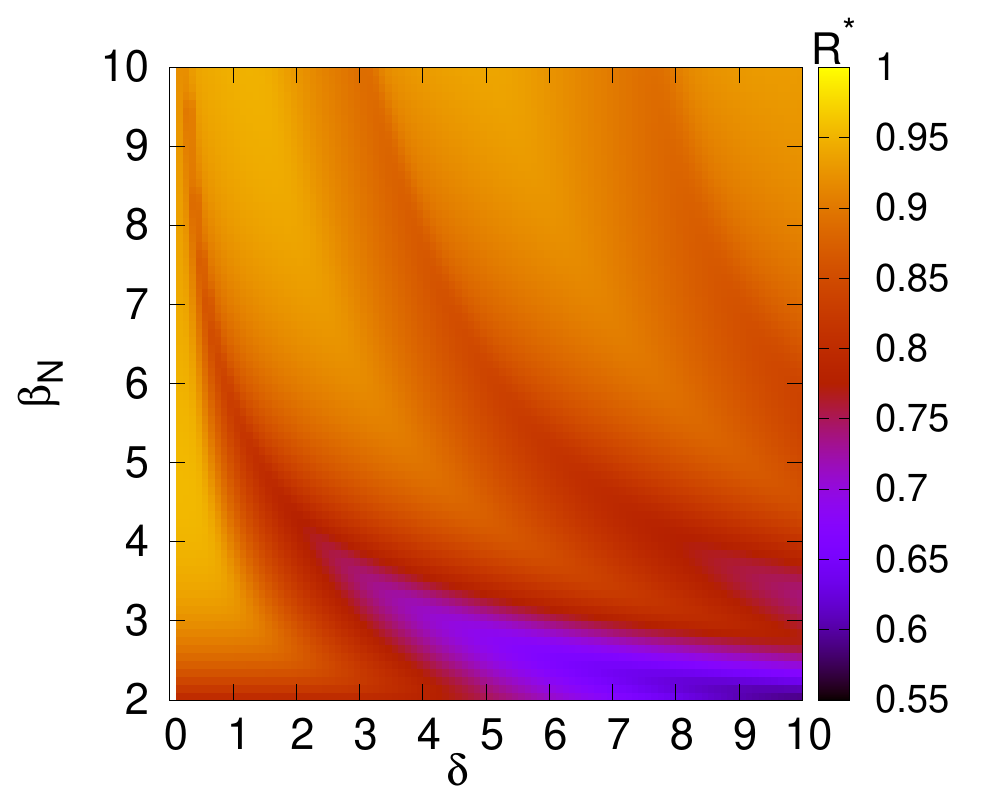}
  \caption{Phase space $\beta_N \times \delta$ for the final epidemic size $R^*$. The oscillations of $R^*$ in relation to both parameters are present for all studied values. Note that the final epidemic size decreases with $\delta$ mainly for low $\beta_N$ values. }
  \label{vardelta}
\end{figure}

Next we present the parameter space $\beta_N \times \delta$ for the final density of removed agents, $R^*$ in Figure \ref{vardelta}. As expected, increased disease risk perceptions leads to a smaller final density of removed agents. Nevertheless, it is clear that this behavior is not trivial, and different infection rates result in large oscillations. It is interesting to note that the valleys and peaks follow, on average, an inverse proportion with $\delta$. For instance, for a fixed value of $R^*$, $\beta_N \propto 1/\delta$. Note that the value of $I_{max}$ is highly dependent on $\delta$ but does not change considerably with $\beta_N$.

\begin{figure*}
\centering
  \includegraphics[width=8cm]{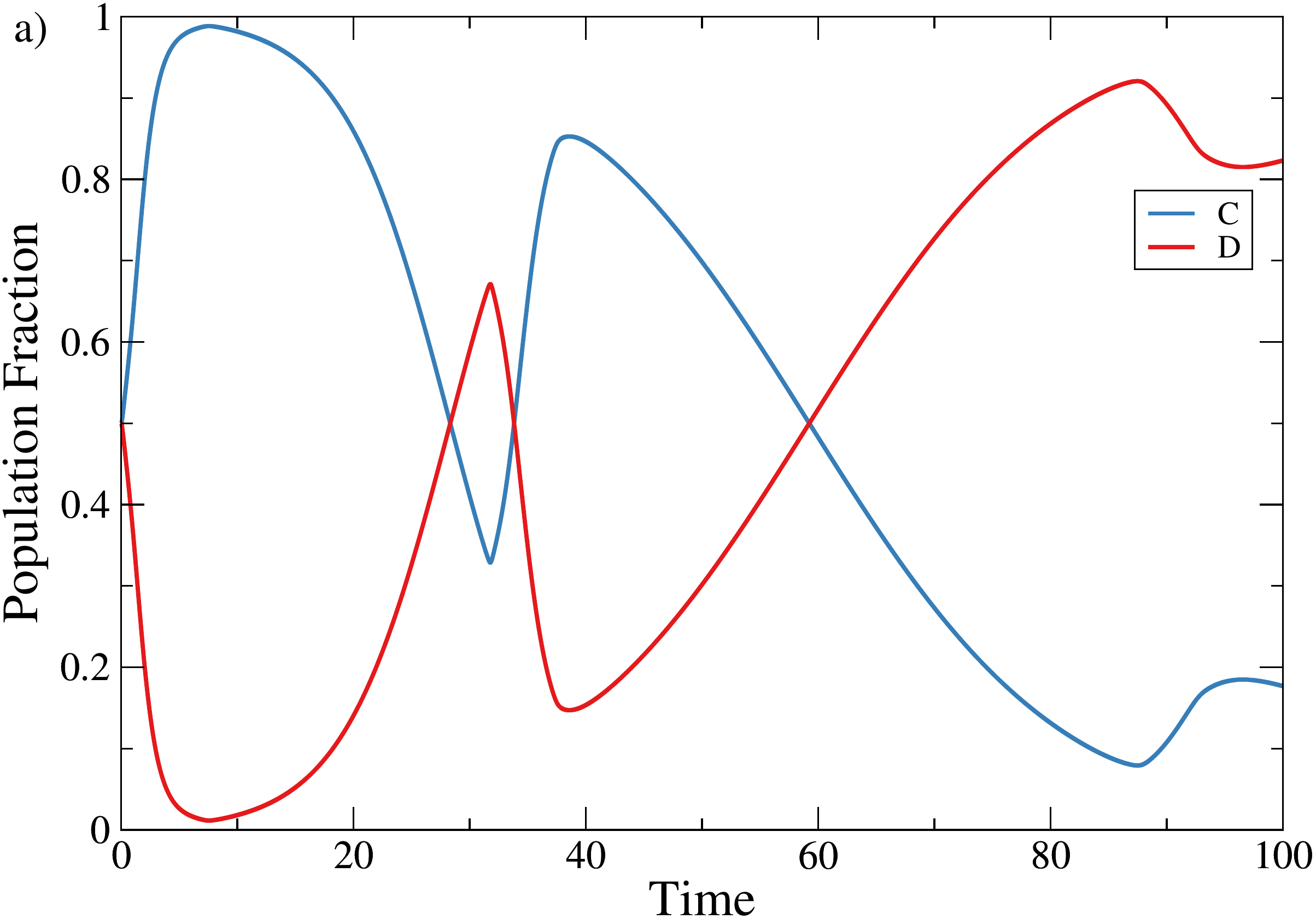} 
  \includegraphics[width=8cm]{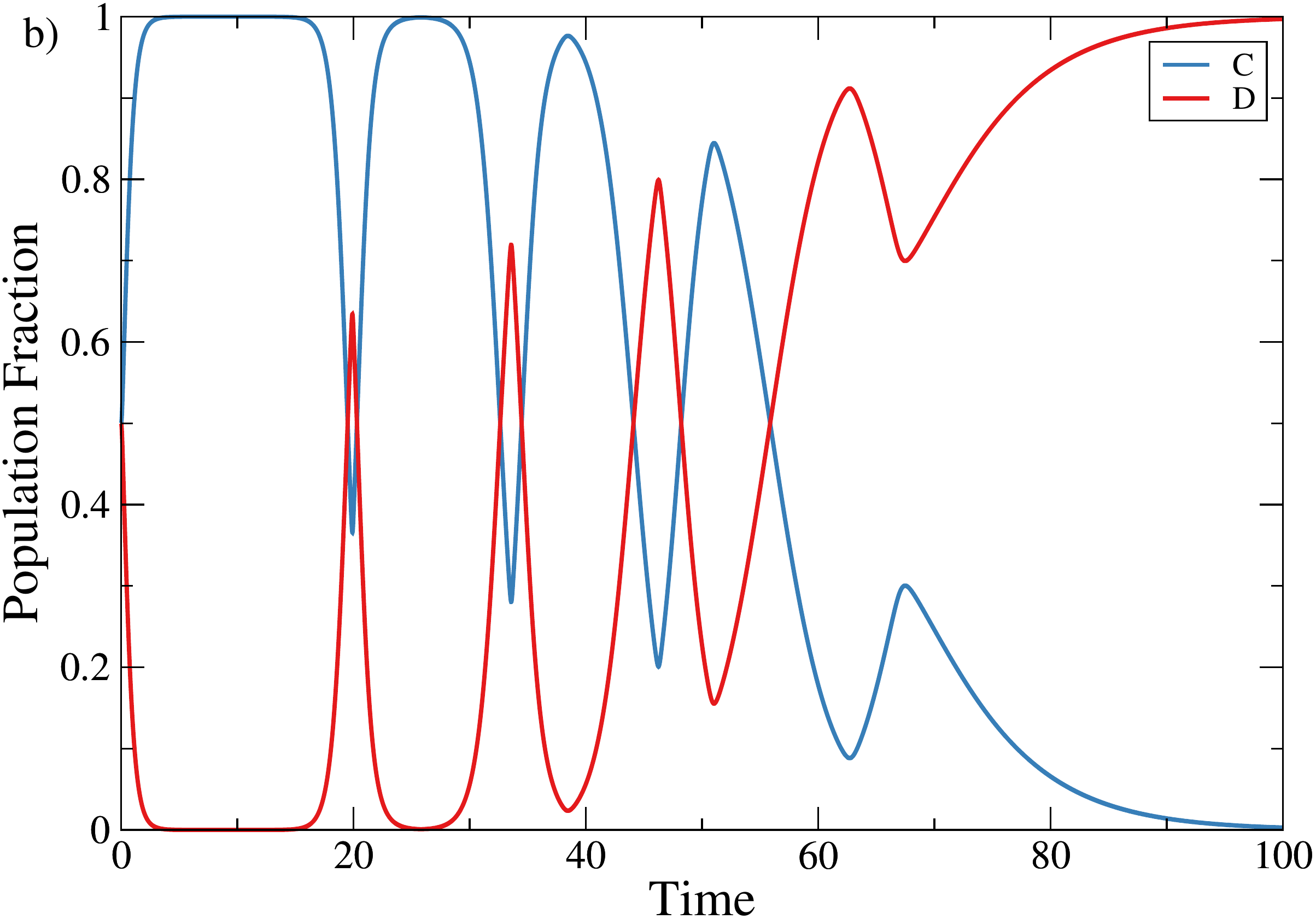}  
  \caption{Strategy adoption evolution for different coupling constant values $\tau$. In a) we present a value corresponding to half the time-scale of the epidemics, i.e. $\tau=0.5$. Figure b) presents a time-scale twice as fast, $\tau=2$. The peaks in the defector fraction always correlates to peaks in the total infected population $I$. Greater $\tau$ values leads to more frequent oscillations in the strategy distribution, and consequently more infection peaks with lower heights. Here we used $\delta=10, \beta_N=10$.}
  \label{tauvar}
\end{figure*}

As $\tau$ is the coupling constant between the epidemic and evolutionary game dynamics, it is correlated with how quickly a population is able to respond to new information regarding the current disease situation. Figure ~\ref{tauvar} reports the effects of different $\tau$ in the evolution of the strategies. Notably, increasing its value causes strategy changes to become more frequent. This in turn entails more oscillations in the whole population. Every peak in the defector density also leads, eventually, to a peak in the density of infected agents, $I$.
Variations in $\tau$ do not change the final infection size considerably. We also note that variations in the irrationality parameter, $k$, did not drastically affect the dynamics for reasonable values ($0.01<k<2$). The main effect of decreasing $k$ is to make the strategy adoption curves sharper around the inflection points. On the other hand, a high irrationality parameter makes the strategy changes more smoothly in time. 

Finally, we generalize the results of the proposed model according to the evolutionary game theory framework. It is a known result that the strategy equilibrium of a classical game is invariant in relation to the multiplication and/or sum of a constant value over all payoffs~\cite{Szabo2007}. Therefore, we can simplify the proposed payoff structure, leaving intact the central characteristics of the game. This allows us to obtain relevant information regarding the general game class.
We first sum $\Omega$ in both payoffs and then divide them by $\beta_N \delta$. Using $\epsilon=\Omega/\beta_N \delta $, we get the simplified version:

\begin{align}\label{general}
\pi_Q&=0. \\
\pi_N&=\epsilon - I. \label{general2}
\end{align}

Note that $\epsilon$ is the ratio between the perceived cost of quarantine and the cost of getting infected. By definition, $0<\epsilon<1$ as we always expect $\Omega < \beta_N \delta$, i.e. the cost of performing a quarantining is smaller than that of being infected. This general payoff structure correctly predicts the most essential feature of our model, i.e. the best strategy is to stay on quarantine if there are many infected agents ($I>\epsilon$), and leave quarantine in the opposite case. This is very similar to the anti-coordination game class, where the best strategy is to do the opposite of your opponent. Here, however, the main factor to consider is the number of infected agents, and not of quarantined ones. 

If everyone is undergoing a quarantine, one has a big incentive to avoid such strategy. On the other hand, if everyone is not taking quarantine precautions, one has a big incentive to do so. This general payoff structure is similar to the free-ride scenario obtained in vaccination games ~\cite{Fu2011, d-onofrio_jtb11} and other models with mitigation policies~\cite{Steinegger2020, Rowlett2020, Chowdhury2020, Kabir2019, reluga_ploscb10, Poletti2009, Schecter}. The inflection point where defection becomes more advantageous can be clearly stated as $I'=\epsilon$.
Differently from a classic game, however, $I=I(t)$, that is, the number of infected agents in our model is time dependent and will depend on the number of agents using the strategy $Q$ or $N$.
Note however that such payoff manipulation only makes the classic game equilibrium invariant, not its evolutionary counterpart. For the population dynamics, the payoff multiplication has the equivalent effect of changing the value of $k$ in the transition probability, (\ref{fermi}), i.e.  $k'=k\beta_N \delta$. 

We also deem important to state how general this payoff structure is. Given that one can re-scale and sum all payoffs by a given constant, the central point of an anti-coordination game is to have incentives that lead players to do the opposite of the majority. 
This can be achieved by the general structure in Eqs. (\ref{general}) and (\ref{general2}), where one strategy has a constant payoff and the other decreases as more agents choose said strategy. This is particularly interesting when looking at mitigation models \cite{Poletti2009, d-onofrio_jtb11, Schecter, Verelst2016} that consider the cooperator payoff as also depending on the number of infected individuals.
For a general case, we could propose that $\pi_Q'=-aI-b$ and $\pi_N'=-cI-d$, with all constants being greater than zero, and $c>a$, as in most mitigation game models \cite{Poletti2009, d-onofrio_jtb11, Schecter, Manfredi2013}. 
We can re-scale such payoff so $\pi_Q'=-b+d~,~\pi_N'=-(c-a)I$. In other words, apart from the constant naming, if we call $b+d=\Omega$ and $c-a= \delta \beta_N$, we get our model back.
In the context of game theory, as long as $\Omega<\delta \beta_N$, the payoff always grows as one chooses the less frequent strategy, maintaining the anti-coordination game.

It is also possible to show that the model is different from the $SIR$ model with two distinct infection rates. Using the definition $S=S_Q+S_N$ and $I=I_Q+I_N$ we see that:
\begin{align}
&\dot{S}=-I_Q(\beta_a S_N +\beta_Q S_Q)-I_N(\beta_N S_N +\beta_a S_Q) \\
&  \dot{I}=I_Q(\beta_a S_N +\beta_Q S_Q)+I_N(\beta_N S_N +\beta_a S_Q)-\gamma I \\
&\dot{R}=\gamma I
\end{align}
Since the flux ($\Phi$) terms regard only transitions between the same epidemiological compartment, they vanish when we look only at the total epidemiological level of the population. Even so, we see that the model does not reduce to the $SIR$ model with two infection rates. Indeed we cannot totally disappear with the sub-population terms.

Furthermore, we can also consider the population at the level of strategy adoption dynamics. $C$ and $D$ represent the density of cooperators and defectors respectively. For the proposed model we have $C=(S_Q+I_Q)/(S+I)$, and since we only have two strategies, $D=1-C$. The rate of change in the strategies comes only from the strategy flux terms $\Phi_S$ and $\Phi_I$. In other words, $\dot{C}=-\Phi_S-\Phi_I$. Using Equations (\ref{phisimpleS}) and (\ref{phisimpleI}), we obtain:
\begin{align}\nonumber
\dot{C}&= (S_N+I_N)(S_Q+I_Q)\Theta(\pi_N,\pi_Q) \\ \nonumber
&- (S_Q+I_Q)(S_N+I_N)\Theta(\pi_Q,\pi_N) 
\end{align}

Re-arranging the terms and noting that $S_Q+I_Q=C(S+I)~,~S_N+I_N=D(S+I)$, and that $S+I=1-R$, we finally obtain:
\begin{eqnarray}
\dot{C}&=&(1-R)^2CD[\Theta(\pi_N,\pi_Q)-\Theta(\pi_Q,\pi_N)]
\end{eqnarray}

The first term, $(1-R)^2$, modulates the speed of the strategy change ($\dot{C}$), as it is related to the total available population allowed to vary the strategies. Most important, however, is the rest of the equation, which is precisely the usual mean-field form of the master equation for the evolution of cooperation in a two strategy game, such as the prisoner's dilemma ~\cite{Szabo2007}.
We can observe that the proposed model is self-consistent and returns the evolutionary game when we only look at the strategy densities. At the same time, (numerically) the model also returns the classic SIR dynamics with two infection rates when we make $\tau=\beta_a=0$, i.e. when we turn off the strategy dynamics and cross infection terms.

\section{Conclusions}

A common approach to analyze complex systems is to isolate its essential elements and features, trying to filter out less relevant components. Such is the case of social behaviors and disease spreading, two intricate processes that, mainly for the sake of simplicity, are often analyzed separately. 
In order to describe their dynamics, identifying their essential elements and interactions, it is fundamental to define a model able to capture, as much as possible, the observed phenomena while maintaining its simplicity.
Due to the relevance of the behavioral component, in particular epidemic situations such as the COVID-19 crisis, here we proposed a theoretical framework devised to combine social strategies with epidemic spreading.
To this end, we present a simplified version of the epidemiological SIR model merged with an evolutionary game that allows agents to rationally choose between a voluntary quarantine or a normal lifestyle during the spreading of a generic disease.
Following this approach, we obtain a single compartmental model that integrates into the same time scale the rational decision making, from game theory, and the epidemiological dynamics of the SIR model. The latter has been chosen as a test case, however, the proposed model can also be realized considering other variations, as the SIS and SEAIR models, as well as other game theory frameworks.
The infection and recovery rates are given by the epidemiological dynamics, while the strategy changes are controlled by the so-called strategy update rules, widely studied in evolutionary game theory. Nevertheless, the infection rates depend on the chosen strategy, whereas the risk perception and payoff of each strategy depend on the number of infected individuals.

We investigate the model through numerical and analytical approaches. Remarkably, the model presents individual reactions to the disease infection level, which can result in secondary infections and the re-emergence of the disease spreading after most of the population dismiss its risk.
In particular, our results revealed multiple infection peaks for higher disease risk perceptions, very similar to the observed behavior of past epidemic cases with voluntary quarantine measures.
The interplay between the contagion and strategy dynamics exhibited a rich behavior. The main parameter that we studied in the model is the perceived disease risk, $\delta$, i.e. a measure of how strongly the population sees the individual cost of being infected. We show that while this parameter has a small effect on the final infection size, it is most important concerning the infection peak size. Notably, the maximum magnitude of the infection peak is found to be inversely proportional to the disease perceive risk $\delta$.

It is worth to emphasize that for no perceived disease risk, agents decide to avoid quarantine and the population quickly suffers from a widespread infection, resulting in a single and huge peak of simultaneously infected agents. 
As recent events related to the global COVID-19 pandemic have shown, the total infection peak is an observable of paramount relevance. In particular, during these critical scenarios, healthcare systems may risk to collapse, due to the possibility that the amount of infected individuals saturates their total capacity~\cite{Zhang2020}. That is one of the reasons why not only the total epidemic size is important, but also the maximum number of simultaneous infections.
In the proposed model, the inclusion of the perceived disease risk makes individuals prone to quarantine for longer times, resulting in a smaller infection peak. As we increase the perceived risk, multiple smaller peaks emerge. This is a direct result of the interconnection between two complex processes, i.e. disease spreading by the SIR model, and rational strategy choices by the evolutionary game dynamics.
We see that for high values of $\delta$, the disease can stay active for longer times and present more infection waves. Nevertheless, those peaks are shorter and the maximum number of simultaneous infections is highly dependent on $\delta$, quickly diminishing as the disease risk perception increases.

We also perform a payoff analysis to find the optimum mixed strategy for a given number of infected individuals. This allows us to analytically obtain the inflection point of the strategy adoption dynamics. This may be used to understand both the dependence of the most used strategy as a function of the infection number, and when the next infection wave can emerge again.
Analyzing other parameters we find that the coupling constant $\tau$ is responsible for changing the speed of the population response to new infections, i.e. how fast the strategy adoption occurs, but has no strong effect on the infection peak size. In the same way, the irrationality parameter $k$ can change the properties of the strategy adoption dynamics without changing its inflection points or the infection peak size.
Lastly, we show that the model is self-consistent and returns the usual replicator equation when looking only at the strategy fractions of the population dynamics. Likewise, when we turn off the interactions between the populations ($\tau=\beta_a=0$) we get back two separated SIR populations, evolving independently.

Overall, the achieved results point to the importance of the disease perceived risk in the spreading dynamics and how such an ingredient can be included in more realistic modeling. 
The area of behavioral epidemiology is relatively recent, and evolutionary game theory and sociophysics seem to have much to add with their approaches. As examples, we cite recent works that have highlighted how evolutionary game dynamics can be used together with an epidemiology-based approach to model social contact behavior such as corruption and rumor spreading ~\cite{Bauza2020, Lu2020a, Askarizadeh2019, Amaral2020b, Lee2019, Capraro2019}. In this sense, we believe that this model can be used as an initial framework to understand more complex phenomena regarding behavioral epidemiology, especially the integration of game theory in compartment models.

\section*{Acknowledgements}
The authors would like to thank the insightful comments made by Eduardo Perovano. This research was supported by the Brazilian Research Agencies CNPq (proc. 428653/2018-9)and Fapemig, project  APQ-01985-18.

\bibliographystyle{elsarticle-num}

\end{document}